\documentclass[12pt,a4paper]{article}
\usepackage[font=footnotesize,labelfont=bf,width=1.1\textwidth]{caption}
\usepackage{authblk}
\usepackage{epsfig,amsmath,amsfonts,amssymb,cancel}
\usepackage{graphicx,latexsym,bbold,mathbbol,color}
\usepackage{stackrel,epstopdf,xcolor,hyperref,cite}
\usepackage[hang,flushmargin]{footmisc}

\newcommand{\be}{\begin{equation}}
\newcommand{\ee}{\end{equation}}
\newcommand{\bea}{\begin{eqnarray}}
\newcommand{\eea}{\end{eqnarray}}
\def\eqa{&=&} 
\def\ccr{\nonumber\\} 
\def\la{\langle}
\def\ra{\rangle}

\begin{document}
\numberwithin{equation}{section}

\title{{\Large\bf On the simplified path integral on spheres}}

\author[a,c,d]{Fiorenzo Bastianelli 
}  
\author[b,c,d]{Olindo Corradini 
}

\affil[a]{{\small\it Dipartimento di Fisica ed Astronomia, Universit{\`a} di Bologna,\protect\\[-2.5mm]
Via Irnerio 46, I-40126 Bologna, Italy}}

\affil[b]{{\small\it Dipartimento di Scienze Fisiche, Informatiche e Matematiche, \protect\\ [-1mm]
Universit\`a degli Studi di Modena e Reggio Emilia,\protect\\[-2.5mm] Via Campi 213/A, I-41125 Modena, Italy}}

\affil[c] {{\small\it INFN, Sezione di Bologna,  Via Irnerio 46, I-40126 Bologna, Italy}}

\affil[d] {{\small\it Max-Planck-Institut f\"ur Gravitationsphysik, Albert-Einstein-Institut, \protect\\[-2mm]
Am M\"uhlenberg 1, D-14476 Golm, Germany}}

\affil[ ]{\protect\\[-1mm]\small {\it Email:} {\tt \href{fiorenzo.bastianelli@bo.infn.it}{fiorenzo.bastianelli@bo.infn.it}, \protect\\[-1mm]\hspace{.075cm}  \href{olindo.corradini@unimore.it}{olindo.corradini@unimore.it}}}

\date{}

\maketitle

\abstract{We have recently studied a simplified version of the path integral for a particle on a sphere, 
and more generally on maximally symmetric spaces, and proved that  Riemann normal coordinates allow
the use of a quadratic kinetic term in the particle action. The emerging  linear sigma model contains 
a scalar effective potential that reproduces the effects of the curvature.
We present here further details on the construction, and extend its perturbative  evaluation to orders high enough
to read off the type-A trace anomalies of a conformal scalar in dimensions $d=14$ and $d=16$.}

\begin{center}
\end{center}


\section{Introduction}
Path integrals for point particles find useful applications in worldline treatments of quantum field theories.
In particular, path integrals for particles on curved spaces allow 
to study gravitationally interacting field theories.
In this paper, after reviewing the simplified path integral for a nonrelativistic particle on a sphere
that has been introduced recently in \cite{Bastianelli:2017wsy}, by presenting further details on its construction, we 
extend its perturbative calculation to orders high enough  to be able to read off the type-A trace anomalies of a 
conformal scalar field in dimensions $d=14$ and $d=16$.

The standard action of a nonrelativistic particle has the form of a nonlinear sigma model in one dimension.
The nonlinearities present in the kinetic term make the definition of the path integral rather subtle, carrying the necessity 
of specifying a regularization scheme together with the fixing of corresponding finite counterterms.
 The latter are needed for specifying a well-defined quantum theory, see   
\cite{deBoer:1995cb, Bastianelli:1998jm, Kleinert:1999aq, Bastianelli:2000nm} for the known regularization schemes. 
The development of those  regularization schemes was
 prompted by the desire of extending the quantum mechanical method of computing chiral
anomalies \cite{AlvarezGaume:1983at, AlvarezGaume:1983ig, Friedan:1983xr}
to trace anomalies \cite{Bastianelli:1991be, Bastianelli:1992ct}. A comprehensive account may be found in the book
\cite{Bastianelli:2006rx}.

A simplified version of the path integral for the case of maximally symmetric spaces, like spheres, has been discussed 
and proved recently in \cite{Bastianelli:2017wsy}. It builds on an old proposal \cite{Guven:1987en} of constructing 
the path integral  by making use of Riemann normal coordinates. These special coordinates are supposed to 
make consistent the replacement of the nonlinear sigma model by a linear one. 
At the same time the inclusion of a suitable effective scalar potential is shown to reproduce the effects of the curvature.  
That this is indeed possible was proved in \cite{Bastianelli:2017wsy} for the case of  maximally symmetric spaces, 
leaving the more difficult question of its validity on arbitrary geometries unsettled.
In the present paper we review the construction on maximally symmetric spaces, and present a detailed perturbative 
evaluation of the path integral, which in particular allows us to identify the trace anomalies 
of a conformal scalar field in dimensions 
$d=14$ and $d=16$. The maximally symmetric background gives information on the so-called type-A trace 
anomaly \cite{Deser:1993yx}, which is proportional to the Euler density of the curved background.
Other methods for identifying the type-A trace anomalies in higher dimensions are probably more efficient, 
see for example \cite{Copeland:1985ua, Cappelli:2000fe, Diaz:2008hy,Dowker:2010qy}, but the path integral construction 
is certainly more flexible, allowing in principle for the calculation of other observables, as exemplified by the
various applications of the worldline formalism (see  \cite{Schubert:2001he} for a review in flat space, and 
\cite{Bastianelli:2002fv, Bastianelli:2002qw, Bastianelli:2005vk, Bastianelli:2005uy, Bastianelli:2004zp, Hollowood:2007ku,
Bastianelli:2008vh, Bastianelli:2008cu, Bastianelli:2012bn, Bastianelli:2013tsa}
for extensions to curved spaces). In any case, we also apply these alternative methods to check our 
final anomaly coefficients.
 
We start  our paper with  Section \ref{sec:2}  where, by using Riemann normal coordinates on maximally 
symmetric spaces, we prove that the Schr\"odinger equation (the heat equation in our euclidean convention) 
for the transition amplitude can be simplified, so to have a corresponding simplified version of the path integral  
that generates its solutions.
In Section \ref{sec:3} we set up the perturbative expansion of the simplified path integral, and proceed to 
evaluate the transition amplitude at coinciding points, as needed for identifying one-loop effective actions
in scalar QFT through worldlines.  In particular, we calculate all the terms that are needed to identify 
the trace anomalies for space-time dimensions $d\leq16$. 
These anomalies are extracted in Section~\ref{sec:4}, and recomputed in Section
\ref{sec:5} with the alternative methods mentioned earlier to show the consistencies of these different approaches. 
Eventually, we present our conclusions and outlook in Section~\ref{sec:6}.

\section{Transition amplitude and path integral on spheres} 
\label{sec:2} 

The classical dynamics of a nonrelativistic particle of unit mass in a curved $d$-dimensional space is described by 
the lagrangian
\be 
 L(x,\dot x)=\frac12 g_{ij}(x)\dot x^i\dot x^j~,
 \label{nonlinear}
 \ee 
where  $g_{ij}(x)$ is the metric in an arbitrary coordinate system and $\dot x^i=\frac{d x^i}{dt}$.
The corresponding hamiltonian reads
\be
H(x,p)= \frac12 g^{ij}(x)p_i p_j ~,
 \ee
where $p_i$ are the momenta conjugated to $x^i$.  Upon canonical quantization the classical hamiltonian identifies
a quantum hamiltonian operator
\be
\hat H_\xi (\hat x,\hat p) = \frac12 g^{-\frac14} (\hat x) \hat p_i g^{\frac12} (\hat x) g^{ij}(\hat x)
\hat p_j g^{-\frac14} (\hat x)
+\frac{\xi}{2} R(\hat x)~,
\ee
where the ordering ambiguities between the $\hat x^i$ and $\hat p_i$ operators have been partially fixed 
by requiring background general coordinate invariance (here $g(x)\equiv\det g_{ij}(x)$). 
The remaining ambiguities are parametrized by  the free coupling constant  $\xi$ that multiplies the scalar 
curvature $R$. Interesting values of this coupling are $\xi=0$ that defines the minimal coupling,
 $\xi=\frac{d-2}{4(d-1)}$ for the conformally invariant coupling in $d$ dimensions, and  $\xi=\frac14$ that allows 
 for a supersymmetrization of the model (it appears in the square of the Dirac operator).
 For simplicity, we will set $\xi=0$ in the following discussion, inserting the nonminimal coupling through a scalar potential, 
 when needed.

We are interested in studying the evolution operator in euclidean time $\beta$ (the heat kernel)
\be
\hat K(\beta)=e^{-\beta \hat H_0}~,
\ee
that satisfies the equation
\bea
 -\frac{\partial \hat K(\beta)}{\partial \beta} \eqa \hat H_0\hat K(\beta)  \label{2.5}
 \\[2mm]
\qquad \hat K(0) \eqa  \mathbb{1} \;.
\eea
It is convenient to use  position eigenstates 
\be \hat x^i|x\ra =x^i|x\ra ~,
\ee
normalized as 
\be
 \la x|x'\ra = {\delta^{(d)}(x-x')\over \sqrt{g(x)}} ~,
 \ee
 so that the resolution of the identity is written as
\be
 \mathbb{1}  = \int d^d x \sqrt{g(x)}\, |x\ra \la x|  \;.
 \ee
 Using them, one recognizes that the wave functions $\psi(x) = \la x| \psi\ra $, corresponding 
 to vectors $|\psi\ra$ of the Hilbert space, are scalars under arbitrary change of coordinates.
 In particular the matrix element of the evolution operator  between these position eigenstates
 gives a transition  amplitude 
 \be 
 K(x, x';\beta) = \la x| e^{-\beta \hat H_0} |x'\ra~,
\ee
  that behaves as a biscalar under arbitrary change of coordinates, i.e. a scalar at both points $x$ and $x'$.  It satisfies the heat equation
  (we use units with $\hbar=1$)
 \bea
-\frac{\partial}{\partial \beta} K(x, x';\beta) \eqa -\frac12 \nabla^2_x \,  K(x, x';\beta)   \label{hk1} \\[2mm]
K(x, x';0) \eqa \frac{\delta^{(d)}(x- x')}{\sqrt{g(x)}}~,
\eea
where $\nabla^2_x$  is the scalar laplacian $\nabla^2 =\frac{1}{\sqrt{g}} \partial_i \sqrt{g} g^{ij} \partial_j $
acting on the $x$ coordinates.
This equation corresponds precisely to the matrix elements of the operatorial equation
\eqref{2.5} between position eigenstates.
 Its solution may be given a well-defined path integral representation 
in terms a nonlinear sigma model action \cite{Bastianelli:2006rx}.

In order to simplify the heat equation and the corresponding
path integral, we  choose to work with Riemann normal coordinates, that are 
reviewed in Appendix \ref{appA}.
 We first transform the transition amplitude  into a bidensity by defining 
\be
\overline K(x,x', \beta) = g^{\frac14}(x) K(x,x', \beta) g^{\frac14}(x') ~,
\ee
so that 
equation \eqref{hk1} takes the form
 \bea
-\frac{\partial}{\partial \beta} \overline K(x, x';\beta) \eqa -\frac12 g^{\frac14}(x) 
\nabla^2_{x}  \Big ( g^{-\frac14}(x) \overline  K(x, x';\beta)  \Big )  \label{hk2} 
\\[2mm]
\overline K(x, x';0) \eqa \delta^{(d)}(x- x')  \;.
\eea
One may evaluate  the differential operator appearing on the right hand side of eq. \eqref{hk2} 
to obtain the identity
\be
 -\frac12 g^{\frac14} \nabla^2  \, g^{-\frac14} = -\frac12 \partial_i g^{ij} \partial_j +V_{eff}~,
 \label{diff-op}
\ee
where in the first addendum derivatives act through, while the effective scalar potential is given by
\be
V_{eff} = -\frac12  g^{-\frac14} \partial_i \sqrt{g} g^{ij} \partial_j g^{-\frac14} ~,
\ee
where all derivatives stop after acting on the last function. The heat equation \eqref{hk2}  now
reads more explicitly as
\be
-\frac{\partial}{\partial \beta} \overline K(x, x' ;\beta) =
\Big (-\frac12 \partial_i  g^{ij}(x) \partial_j +V_{eff}(x)  \Big )\overline K(x, x';\beta) \;.
\label{hk3}
\ee
At this stage we are ready to use 
the properties of Riemann normal coordinates, centered at the point $x'$, 
to show that this heat equation simplifies further to 
\be
-\frac{\partial}{\partial \beta} \overline K(x, x';\beta) =
\Big (-\frac12 \delta^{ij}\partial_i \partial_j +V_{eff}(x)  \Big )\overline K(x, x';\beta) ~,
\label{hk4}
\ee
as on maximally symmetric space, in Riemann normal coordinates,
one may replace the metric $g^{ij}(x)$ appearing in  the term $\partial_i g^{ij}(x) \partial_j $ 
by the constant metric $\delta^{ij}$. Note that the heat kernel equation \eqref{hk4}
contains now an hamiltonian operator 
\be
H=-\frac12 \delta^{ij}\partial_i \partial_j +V_{eff}(x)  
\ee
which is interpreted as that of a particle on a flat space (in cartesian coordinates) interacting with an effective 
scalar potential $V_{eff}$ of quantum origin (it would be proportional to $\hbar^2$ in arbitrary units).
 
For the replacement of  $g^{ij}(x)$ with $\delta^{ij}$ to be valid, one must show that
\begin{align}
\Big(\partial_i g^{ij}(x) \partial_j-\delta^{ij}\partial_i\partial_j\Big) \overline K(x, x';\beta) =0~.
\label{Diff}
\end{align}    
To see this, we recall that $x'=0$ is the chosen origin of the Riemann normal coordinates,  
and using the inverse metric given in \eqref{A8} we find that the equation that we must verify takes the form
\begin{align}
\Big( h(x) P^{ij}(x)\partial_i\partial_j +\partial_i\big(h(x)P^{ij}(x)\big)\partial_j\Big) \overline K(x, 0;\beta) =0 ~,
\label{Diff-1}
\end{align}   
where the projector $P^{ij}(x)$ and the function $h(x)$  are given by 
\bea
P^{ij}(x) \eqa    \delta_{ij} - \frac{ x_i x_j}{x^2}  
\\[2mm] 
h(x) \eqa -\frac{f(x)}{1+f(x)} =\frac{2 (Mx)^2}{1- \cos(2 Mx)} -1~,
\eea
 as discussed in Appendix \ref{appA}.
The function $h(x)$ is a function of only $x^2=\delta_{ij}x^i x^j$, 
since it is even in $x\equiv\sqrt{\delta_{ij}x^i x^j}$. This is a consequence of the maximal symmetry of the sphere.
The explicit evaluation of the derivatives  appearing in \eqref{Diff-1} produces 
(recalling the orthogonality condition $P^{ij}x_j=0$)
\bea
h(x) P^{ij}(x)\partial_i\partial_j \overline K(x, 0;\beta)
\eqa
2h(x) \delta_{ij} P^{ij}(x)\frac{\partial}{\partial x^2}\overline K(x, 0;\beta) \ccr[2mm]
\eqa 2(d-1) h(x)\frac{\partial}{\partial x^2}\overline K(x, 0;\beta) ~,
\eea
and 
\begin{align}
\partial_i\big(h(x)P^{ij}(x)\big)\partial_j\overline K(x, 0;\beta) =-2(d-1) h(x)\frac{\partial}{\partial x^2}\overline K(x, 0;\beta)\;.
\end{align} 
The two terms cancel each other, so that we have indeed verified eq. \eqref{Diff-1} and the correctness of the 
heat kernel equation \eqref{hk4} for our problem.

To summarize,  we are led to consider the  euclidean Schr\"odinger equation  
\be
-\frac{\partial}{\partial \beta} \overline K(x, x'\beta) =
\Big (-\frac12 \delta^{ij}\partial_i \partial_j +V_{eff}(x)  \Big )\overline K(x, x';\beta) ~,
\ee
valid in Riemann normal coordinates centered at $x'$,
to describe the quantum motion of a particle on a sphere.
This equation can now be solved by a standard path integral 
\be
\overline K(x, x';\beta) = \int_{x(0)=x'}^{x(\beta)=x} Dx\ e^{ - S[x]}  ~,
\label{simple-pi}
\ee
where the action is that of a linear sigma model augmented by an effective potential 
\be
S[x] =  \int_0^\beta  \!\!\! dt   \left (  \frac{1}{2} \delta_{ij}\dot x^i \dot x^j +  V_{eff}(x)  \right ) \;.
\label{linear}
\ee
The required effective potential is
\bea
V_{eff}(x) \eqa  -\frac12  g^{-\frac14} \partial_i \sqrt{g} g^{ij} \partial_j g^{-\frac14} 
\eea
and can be computed in terms of the function $f(x)$ given in \eqref{A4} and \eqref{A5} as 
\be
 V_{eff}(x) =   \frac{(d-1)}{8}  \Biggl[  \frac{(d-5)}{4}  \left (\frac{f'(x)}{1+f(x)}\right)^2
 +  \frac{1}{1+f(x)} \left ( \frac{(d-1)}{x} f'(x) + f''(x) \right) \Biggr]
  \ee
 or, more explicitly, as 
\be
V_{eff}(x)= \frac{d(1-d)}{12} M^2
+ \frac{( d-1) (d-3)}{48}  \frac{\Big(5 (Mx)^2-3 +\left((Mx)^2+3\right) \cos (2 M x)\Big)}{x^2 \sin ^2 (Mx)}  \;.
\label{eff-pot}
   \ee
\begin{figure}[h!]
\begin{center}
\includegraphics[scale=.375]{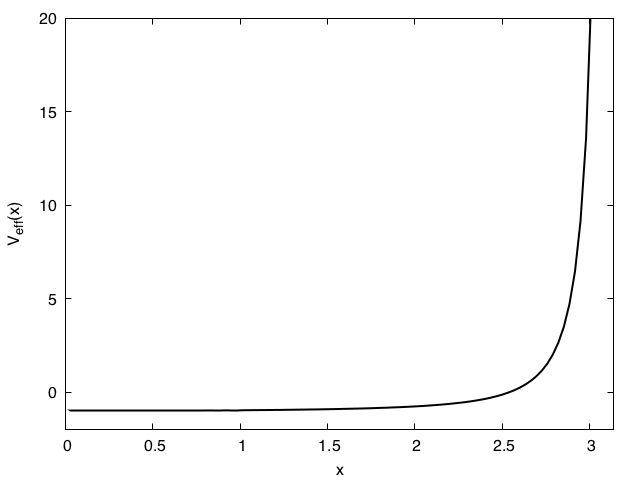}
\end{center}
\caption{Graphical representation of the effective potential for $d=4$ and $M=1$ (sphere of unit radius).\label{fig:pot}}
\end{figure}
The point $x=0$ is the origin of the Riemann normal coordinates (say the north pole), and we see that the effective 
potential becomes singular when $x=\pi/M$, i.e. at the south pole. We plot the radial behavior of this potential in
Figure~\ref{fig:pot}.
Note that the potential is basically flat around $x=0$, but diverges at $x=\pi/M$ that corresponds to the south pole of the sphere, and which is a coordinate singularity of the patch considered (the so-called normal neighborhood).

\section{Perturbative expansion} 
\label{sec:3}

The path integral expression  for the transition amplitude on a sphere 
in terms of the linear sigma model \eqref{linear}  is much simpler than the corresponding one 
with the nonlinear sigma model with lagrangian \eqref{nonlinear}.
In particular, its perturbative evaluation is straightforward as no perturbative vertices are produced from the kinetic 
term and from the path integral measure, which is translational  invariant. 
The only vertices are those without derivatives arising from the expansion of $V_{eff}$.
They produce Feynman graphs that do not need any regularization.

Let us now describe the perturbative expansion of the transition amplitude \eqref{simple-pi} by considering 
coinciding initial and final points, $x'=x$, which give the diagonal part of the heat kernel 
$\overline K(x, x;\beta)$. This is enough to identify  one-loop effective actions and anomalies in QFT using
worldlines. We must use Riemann normal coordinates centered at $x'=x$, and
in such coordinates the diagonal heat  kernel evaluated at the origin is denoted by $\overline K(0, 0;\beta)$.

To start with we rescale the time to $\tau =\frac{t}{\beta}$ to write the action in the form
  \bea
S[x] = \int_{0}^{1} \!\!\! d\tau  \left (  \frac{1}{2\beta} \delta_{ij}\dot x^i \dot x^j + \beta  V_{eff}(x) \right )~,
\eea
which shows that, in an expansion for short times $\beta$, the leading behavior is due to the 
kinetic term, while the effective potential $V_{eff}$ gives perturbative corrections.
The perturbative expansion of the path integral is obtained by setting
\be
S[x] = S_{free}[x] +S_{int}[x]~,
\ee
with
 \bea
S_{free}[x]\eqa \frac{1}{\beta} \int_{0}^{1} \!\!\! d\tau\,  \frac{1}{2} \delta_{ij}\dot x^i \dot x^j  
\label{act1}
\\[2mm]
S_{int}[x] \eqa \beta  \int_{0}^{1} \!\!\! d\tau\, V_{eff}(x) ~,
\label{act2}
\eea
so that  the transition amplitude at coinciding points $x = x'= 0$ in Riemann normal coordinates
may be written as
\be
\overline K(0, 0;\beta)  =\frac{\langle e^{- S_{int}} \rangle }{(2\pi \beta)^{d\over 2}}  ~,
\label{pert-exp}
\ee
where $\la ...\ra$ denotes a normalized correlation function with the free path integral weight.
The normalization is just the one of a free particle and corresponds to the exact path integral performed with
$S_{free}$.

The free propagator for the dynamical variables $x^i(\tau)$, vanishing both at $\tau=0$ and $\tau=1$ 
(Dirichlet boundary conditions with initial and final point fixed at the origin of the Riemann coordinates), 
is obtained by inverting the differential operator of the kinetic term  in \eqref{act1}   and reads
\be
\la x^i(\tau)x^j(\sigma) \ra =-\beta \delta^{ij} [\, \partial^2_\tau \, ]_{(\tau, \sigma)}^{-1} =-\beta \delta^{ij} \Delta(\tau,\sigma)~,
\ee
where the Green function $\Delta (\tau,\sigma)$ with vanishing  Dirichlet boundary conditions is given by
\bea
\Delta (\tau,\sigma)   \eqa   (\tau-1)\sigma\, \theta(\tau-\sigma)+(\sigma-1)\tau\, \theta(\sigma-\tau) \ccr
\eqa \frac12 |\tau-\sigma| -\frac12(\tau+\sigma) +\tau\sigma~, 
\label{2.prop}
\eea
where $\theta(x)$ is the Heaviside step function (the regulated value $\theta(0)=\frac12$
is not  needed in the evaluation of the perturbative corrections).
The two expressions are equivalent, and one may use the preferred one. 
The Green function $\Delta (\tau,\sigma)$ satisfies the defining equation
\be
\partial^2_\tau \Delta (\tau,\sigma) = \delta(\tau-\sigma)~,
\ee
and the boundary conditions
\be
\Delta (0,\sigma) = \Delta (\tau,0) =0\;.
\ee

We are now ready to evaluate perturbative corrections.
They are obtained by expanding the effective potential $V_{eff}$, and computing the perturbative terms 
with an application of the Wick theorem (i.e. calculating gaussian averages).
Taylor expanding the potential \eqref{eff-pot} about $x=0$  produces
       \bea 
  V_{eff}(x)\eqa  
M^2 \frac{d(1-d)}{12} + M^2( d-1) (d-3)
\left( \frac{(Mx)^2}{120}  +\frac{(Mx)^4}{756}+\frac{(Mx)^6}{5400}+\right.
\ccr
&+& \left.  \frac{(Mx)^8}{41580}+ \frac{691 (Mx)^{10}}{232186500}+
\frac{(Mx)^{12}}{2806650}+
\frac{3617 (Mx)^{14}}{86837751000} +
O\left(x^{16}\right) \right)~, 
\label{K-coeff}
   \eea
 and  the interaction vertices arising from it may be written as
\be
S_{int}= \beta  \int_{0}^{1} \!\!\! d\tau\,  V_{eff}(x) =
\sum_{m=0}^\infty \ S_{2m}~,
\ee
where $S_{2m}$ is the term containing  the power $(x^2)^m$, with $x^2=\vec{x}^{\, 2} =x^ix_i$.
Their structure is of the form
\be
S_{2m} = \beta  M^{2+2m} k_{2m}  \int_{0}^{1} \!\! d\tau\, (x^2)^m  ~,
\ee
where the overall power of $M^2$ has been factored out,  while the remaining 
numerical coefficients $k_{2m}$ can be read off from \eqref{K-coeff}.

The first term 
\be
S_0 = \beta M^2 \frac{d(1-d)}{12} 
 \ee
 is just a constant, and can be immediately extracted out of  \eqref{pert-exp}
 to give
 \be
\overline K(0, 0;\beta)  =\frac{ e^{- S_{0} +  \cdots}}{(2\pi \beta)^{d\over 2}}      
\ee
also written more explicitly in terms of the scalar curvature $R$ as
 \bea 
\overline K(0, 0;\beta)  
\eqa {1\over (2\pi \beta)^{d\over 2}}  \exp \biggl [ \frac{\beta M^2}{12}d(d-1) +\cdots \biggr ] 
\ccr
\eqa {1\over (2\pi \beta)^{d\over 2}}   \exp \biggl  [\frac{\beta R}{12}  +\cdots  \biggr] \;.
\eea
Notice that this result is exact for the three-sphere $S^3$, as for $d=3$ the remaining part of the effective 
potential vanishes. This answer was obtained long ago by Schulman \cite{Schulman:1968yv},
who used the fact that $S^3$ coincides with the group manifold of $SU(2)$.

The next correction is the first nontrivial one, and arises form the vertex  $S_2$.
Expanding the interaction term in \eqref{pert-exp} as
\be
\langle e^{- S_{int}} \rangle  =  e^{- S_0} (1-  \langle S_2 \rangle + \cdots)
\ee
shows that one must compute the correlation function $\la S_2\ra$. It identifies a loop graph of the form
\begin{center}
\includegraphics[scale=.5]{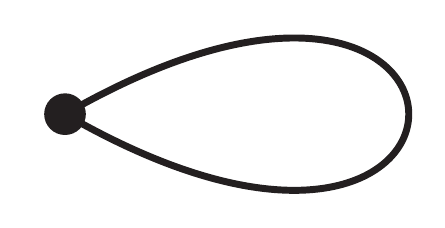}
\end{center}
where the coefficient of the quadratic vertex is found from \eqref{K-coeff},
 and the  propagator is the one given in \eqref{2.prop}. Its calculation proceeds as follows
\bea
 - \langle S_2 \rangle \eqa -\beta M^4 k_2 \int_0^1 \!\! d\tau \, \delta_{ij} \la x^i(\tau) x^j(\tau)\ra
 =  \beta^2 M^4  k_2\, d 
 \int_0^1 \!\! d\tau \, \Delta (\tau,\tau)  \ccr
 \eqa -\frac{\beta^2 M^4}{720} d (d-1) (d-3)~,
 \eea
as the coupling $k_2$ obtained from \eqref{K-coeff} is $k_2= \frac{(d-1)(d-3)}{120}$, 
 while  the integral of the two-point correlation function at coinciding points ($\sigma=\tau$) gives 
 \be 
 \int_0^1 \!\! d\tau \, \Delta (\tau,\tau) =\int_0^1 \!\! d\tau \, (\tau^2-\tau) = -\frac{1}{6} \;.
 \ee
This result can be exponentiated to account for the disconnected contributions arising  from such graphs
at higher orders. Thus, at this perturbative level, we find the transition amplitude 
\be
\overline K(0, 0;\beta)  =\frac{ e^{- S_{0}} }{(2\pi \beta)^{d\over 2}}  e^{- \la S_2 \ra + O(\beta^3)}~,
\ee
 which takes the explicit form 
 \bea 
\overline K(0, 0;\beta)  
\eqa {1\over (2\pi \beta)^{d\over 2}}  \exp \biggl [ \frac{\beta M^2}{12} d(d-1)
-\frac{\beta^2 M^4}{720} d (d-1) (d-3)  +\cdots \biggr ] 
\ccr
\eqa {1\over (2\pi \beta)^{d\over 2}}   \exp \biggl  [\frac{\beta R}{12}   
- \frac{(\beta R)^2}{6!}   {(d-3)\over d (d-1)} +\cdots 
\biggr]  \;.
\eea

In a similar way one may proceed to higher orders.
It is clear that all perturbative corrections appear as powers of $ \beta M^2$, or equivalently $ \beta R$,
as verified by  power counting. In this section we wish to reach order $\beta^8$, so that we must compute
\bea
&&\overline K(0, 0;\beta)  = \frac{e^{-S_0} }{(2\pi \beta)^{d\over 2}}  
\exp \biggl [ - \langle S_2 \rangle
-\underbrace{\langle S_4 \rangle}_{O(\beta^{3})} 
\ccr
&&
\underbrace{ - \langle S_6 \rangle +{1\over 2}  \langle S_2^2 \rangle_c}_ {O(\beta^{4})}
\ccr
&& 
\underbrace{- \langle S_8 \rangle + \langle S_4 S_2 \rangle_c }_ {O(\beta^{5})}
\ccr
&& 
\underbrace{ - \langle S_{10} \rangle + \langle S_6 S_2 \rangle_c  
+ {1\over 2}  \langle S_4^2 \rangle_c  - {1\over 3!}  \langle S_2^3 \rangle_c}_{O(\beta^{6})}
\ccr
&&
 \underbrace{- \la S_{12} \ra   +
\la S_8 S_2\ra_c  +
\la S_6 S_4\ra_c  -\frac12
\la S_4 S_2^2 \ra_c}_{O(\beta^{7})}
\ccr
&&
 \underbrace{-\la S_{14}\ra  
+\la S_{10} S_2\ra_c 
+\la S_8 S_4\ra_c 
+\frac12 \la S_6^2 \ra_c 
-\frac12 \la S_4^2 S_2\ra_c 
-\frac12 \la S_6 S^2_2\ra_c 
+\frac{1}{4!} \la S_2^4\ra_c}_{O(\beta^{8})}
\ccr
&&
+ O(\beta^{9}) \biggr ] \;.
\label{peta}
\eea
The calculation up to $O(\beta^{6})$ was sketched  in \cite{Bastianelli:2017wsy}.
Here we continue through order  $O(\beta^{7})$ and  $O(\beta^{8})$.
As indicated by the notation $\la ...\ra_c$,  it is enough to compute  connected correlation functions only,
as the disconnected pieces have been automatically included by exponentiation.
We report the detailed calculations  in Appendix~\ref{appB}.

Adding all contributions, we summarize our final result for the heat kernel at coinciding points 
\bea 
&&\overline K(0, 0;\beta)  =
{1\over (2\pi \beta)^{d\over 2}} 
 \exp \biggl [ d(d-1)\frac{\beta M^2 } {12}  +d(d-1)(d-3)\biggl( 
 - \frac{(\beta M^2)^2}{720}
 \ccr
 &&
 -\frac{(\beta M^2)^3}{7!} \frac{2(d+2)}{9}
  \ccr
 &&
- {(\beta M^2)^4\over 7!} \frac{(d^2+20d+15)}{360}
\ccr
&& + \frac{(\beta M^2)^5}{11!} \frac{8 (d+2)(d^2-12d -9)}{3}
\ccr
&& 
+ \frac{(\beta M^2)^6}{13!} \frac{8 (1623 d^4 - 716 d^3  - 65930 d^2 - 123572 d  -60165)}{ 315}
   \ccr
   &&
+ \frac{(\beta M^2)^7}{13!} \frac{16 (d+2) (33 d^4 +404 d^3  - 2510 d^2 - 6612 d  -3915)}{315}
   \ccr
  &&
- \frac{(\beta M^2)^8}{17!} \frac{8} {45}
   (12405 d^6 -810668 d^5 -1953995 d^4+17853784 d^3+71217159d^2  \ccr
  &&
+92279700 d  +40157775)
 + O(\beta^{9}) 
 \biggl) \biggr] ~,
 \label{final-1}
\eea
which we present also in terms of the scalar curvature $R$
(recall that $R= M^2 d(d-1)$ with $M=\frac{1}{a}$ the inverse sphere radius) 
\bea 
&&\overline K(0, 0;\beta)  =
{1\over (2\pi \beta)^{d\over 2}}   \exp \biggl [\frac{\beta R}{12} 
- \frac{(\beta R)^2}{6!}   {(d-3)\over d (d-1)} 
- {(\beta R)^3\over 9!} {16 (d-3)(d+2)\over d^2 (d-1)^2} 
\ccr
&&
- {(\beta R)^4\over 10!} {2 (d-3)(d^2+20d+15)\over d^3 (d-1)^3} 
\ccr
&&
+ {(\beta R)^5\over 11!} {8 (d-3)(d+2)(d^2-12d -9)\over 3 d^4 (d-1)^4} 
\ccr
&&
+ \frac{(\beta R)^6}{13!} \frac{8 (d-3) (1623 d^4 - 716 d^3  - 65930 d^2 - 123572 d  -60165)}{ 315 d^5 (d-1)^5 }
   \ccr
   &&
+ \frac{(\beta R)^7}{14!} \frac{32 (d-3) (d+2) (33 d^4 +404 d^3  - 2510 d^2 - 6612 d  -3915)}{45 d^6 (d-1)^6 }
   \ccr
  && 
- \frac{(\beta R)^8}{17!} \frac{8 (d-3)} {45 d^7 (d-1)^7} \Big (12405 d^6 -810668 d^5 -1953995 d^4+17853784 d^3 \ccr
&& \quad +71217159d^2 +92279700 d  +40157775\Big )
 + O(\beta^{9})  \biggr] \;.
 \label{final-2}
\eea

This exponential  can be expanded keeping terms up to order  $O(\beta^{8})$ included,
to read off the heat kernel coefficients at coinciding points $a_n(0,0)$
for the integer $n$ up to $n=8$, defined by
\be
\overline K(0, 0;\beta)  =
{1\over (2\pi \beta)^{d\over 2}}   \sum_{n=0}^\infty a_n(0,0) \beta^n \;.
\label{hk-coeff}
\ee
We will do this in the next section for conformal hamiltonians to extract the so-called type-A trace anomalies.

\section{The type-A trace anomalies}
\label{sec:4} 

One may use the path integral calculation of the transition amplitude on a sphere
to evaluate the type-A trace anomalies of a conformal scalar field.
We have performed this exercise in \cite{Bastianelli:2017wsy} to test the correctness and usefulness 
of the linear sigma model approach. We are now ready to extend those results
to identify the trace anomalies in 
 $d=14$ and $d=16$ dimensions. As reviewed in \cite{Bastianelli:2017wsy}, 
 the trace anomaly of the conformal scalar field
can be related to the transition amplitude of a particle in a curved space by 
\begin{align}
\big\langle T^\mu{}_\mu(x)\big\rangle_{QFT} = \lim_{\beta\to 0} K_\xi(x,x;\beta)~,
\label{trace-an}
\end{align}
where on the left hand side $T^\mu{}_\mu(x)$ is the trace of the  stress tensor of the conformal scalar  
in a curved background, and the expectation value is performed in the corresponding 
quantum field theory. 
The right hand side can instead be viewed as the anomalous contribution arising from the
QFT path integral measure regulated {\it \`a la} Fujikawa \cite{Fujikawa:1980vr}.
The regulator $H_\xi$  corresponds to the kinetic term  of the scalar quantum field theory, which is proportional 
to the conformal laplacian. This is identified with the quantum hamiltonian $H_\xi$ of a particle in a curved space
\be
H_\xi= \frac12 (-\nabla^2 + \xi R) ~, \qquad \xi=\frac{(d-2)}{4(d-1)}~,
\label{conf-ham}
\ee 
which must be
used in evaluating the transition element $K_\xi(x,x;\beta)$ at coinciding points
\cite{Bastianelli:1991be, Bastianelli:1992ct}. It  is understood that 
the $\beta \to 0$ limit  in \eqref{trace-an} picks up just the $\beta$-independent term, as divergent terms are removed 
by the QFT renormalization.  This procedure selects the appropriate heat kernel coefficient $a_n(x,x)$ sitting 
in the expansion of  $K_\xi(x,x;\beta)$, as in  \eqref{hk-coeff}. 
To reproduce the correct conformal hamiltonian \eqref{conf-ham}
we must add a nonminimal coupling through an additional constant potential
\be
V_\xi = \frac12 \xi R~,
\ee
so that we must shift $V_{eff} \to V_{eff} +V_\xi$ in \eqref{act2}. Its effect is to replace
the leading term of  \eqref{final-2} by
\be
{1\over (2\pi \beta)^{d\over 2}}   \exp \biggl  [\frac{\beta R}{12}  +\cdots  \biggr]  
\quad \to \quad
{1\over (2\pi \beta)^{d\over 2}}   \exp \biggl  [\frac{\beta}{12}\left (1 -6 \xi \right) R  +\cdots  \biggr]  
\ee
to obtain the desired amplitude $K_\xi(x,x;\beta)$. Expanding $K_\xi(x,x;\beta)$ at the required order we find 
the trace anomalies in $d$ dimensions
\begin{align}
\big\langle T^\mu{}_\mu(x)\big\rangle_{QFT} = \frac{a_\frac{d}{2}(x,x)}{(2 \pi)^\frac{d}{2}}~,
\label{trace-an1}
\end{align}
which we list in Table \ref{table}, expressing the results also in terms of the sphere radius $a=\frac{1}{M}$.
Of course, one may use Riemann normal coordinates centered at $x$, so that $\sqrt{g(x)}=1$ and
the result in \eqref{final-2} is directly applicable.

{
\renewcommand{\arraystretch}{2}
\begin{table}[h!]
\begin{center}
\begin{tabular}{ | l  | p{7cm} |    p{4cm} | }
    \hline 
  $d$  &   $\la T^\mu{}_\mu\ra$ &  $\la T^\mu{}_\mu\ra$  \\   \hline
2 & \large $\frac{R}{24\, \pi} $ 
& \large $\frac{1}{12\, \pi  a^2 }  $ \\ \hline
4 & \large $-\frac{R^2}{34\, 560\, \pi^2} $ 
& \large $-\frac{1}{240\, \pi^2 a^4}  $ \\ \hline
6 & \large $\frac{R^3}{21\,772\, 800\, \pi^3} $  
& \large $\frac{5}{4\, 032\, \pi^3 a^6} $  \\ \hline
8 & \large  $-\frac{23\, R^4}{339\, 880\, 181\, 760\, \pi^4} $   
&\large  $-\frac{23}{34\, 560\, \pi^4 a^8}$   \\ \hline
10 & \large  $\frac{263\, R^5}{2\, 993\, 075\, 712\, 000\, 000\, \pi^5} $   
& \large $\frac{263}{506\, 880\, \pi^5 a^{10}}  $   \\ 
\hline
12 & \large $-\frac{133\, 787 \, R^6}{1\, 330\, 910\, 037\, 208\, 675\, 123\, 200\, \pi^6}  $
& \large $-\frac{133\, 787 }{251\, 596\, 800\, \pi^6 a^{12}} $\\
\hline
14 & \large $\frac{157\, 009 \, R^7}{1\, 536\, 182\, 179\, 466\, 286\, 307\, 737\, 600\,
\pi^7}  $
& \large $\frac{157\,009 }{232\,243\,200\, \pi^7 a^{14}} $
\\
\hline
16 & \large $-\frac{16\, 215\, 071 \, R^8}{173\, 836\, 853\, 795\, 629\, 301\, 760\, 000\, 000\, 000\,
\pi^8} $
& \large  $-\frac{16\, 215\, 071}{15\, 792\, 537\, 600 \, \pi^8 a^{16}} $
\\
\hline
    \end{tabular}
    \caption{Type-A trace anomalies of a conformal scalar field.
~\label{table}}
    \end{center}
    \end{table}
}

\section{Alternative methods and checks}
\label{sec:5} 
In the present section we check our results on the type-A trace anomalies of a conformal scalar 
by using alternative approaches based on the $\zeta$-function regularization. One method  was used in~\cite{Copeland:1985ua} and later re-elaborated in~\cite{Cappelli:2000fe}. Following the prescription reported in those references, one finds that the type-A trace anomaly on a $d$-sphere is given by
\begin{align}
\left\langle T^\mu{}_\mu\right\rangle = \frac{\Gamma\left( \frac{d+1}{2}\right)}{2\pi^{\frac{d+1}{2}} a^d}\, \zeta_{Y_d}(0)~,
\label{eq:tr-z}
\end{align} 
where $\zeta_{Y_d}(s)$ is the $\zeta$-function associated to the (eigenvalues of the) kinetic operator $Y_d$ of 
the conformally-coupled scalar field on the sphere
\be
Y_d= -\nabla^2 +\frac{(d-2)}{4(d-1)} R
\ee
often called ``Yamabe operator" in the mathematical literature.
Its analytic continuation at $s\to 0$ is given by~\footnote{The expression below coincides with eq. (2.29) of~\cite{Copeland:1985ua},  thanks to the identity ${\displaystyle\sum_{p=0}^{(d-2)/2}}{C_p(d)}\, 2^{-2p}=0$, which is satisfied by the values of $C_p(d)$ of Table~\ref{table2}, as one may check.} 
\begin{align}
\zeta_{Y_d}(0) = \frac{1}{(d-1)!}\sum_{p=0}^{(d-2)/2}\frac{C_p(d)}{p+1}\biggl\{ \left(1-2^{-(2p+1)}\right) B_{2p+2}-2^{-2p}\left(\frac12 p+\frac14\right)\biggr\}~,
\label{eq:z}
\end{align} 
where $B_{2p+2}$ are Bernoulli numbers.  The set of numerical coefficients $C_p(d)$ is the solution of the linear system
\begin{align}
\frac{(n+d-2)!}{n!} =\sum_{p=0}^{(d-2)/2} C_p(d) \left(n+\frac{d-1}{2}\right)^{2p}\,, \quad n=1,\dots, \frac{d}{2}~.
\label{eq:lin-sys}
\end{align}
For $d=14,16$ they read
\begin{table}[h!]
\begin{align*}
\renewcommand{\arraystretch}{2}
\begin{array}{c|c|c|c|c|c|c| c|c}
C_p(d)\, & 0 & 1 & 2 & 3 & 4 & 5 & 6 & 7\\[1mm] \hline
 d=14\, & \frac{108056025}{4096} &-\frac{64408383}{512}&\frac{21967231}{256}&-\frac{308737}{16}&\frac{28743}{16}& -
\frac{143}{2}&\ 1\ &\ \ast \ \\[1mm] \hline
d=16\, &\, -\frac{18261468225}{16384}\, &\,\frac{21878089479}{4096}\, &\, -\frac{3841278805}{1024}\,&\,\frac{230673443}{256}\,&\, -\frac{6092515}{64}\, &\, \frac{77077}{16}\, &\, -
\frac{455}{4}\, &\ 1\ 
\end{array}
\end{align*}
\caption{Solutions of the linear system~\eqref{eq:lin-sys}, for $d=14,16$.\label{table2}}
\end{table}

\noindent
which, inserted into~\eqref{eq:z}, yield 
\begin{align}
\renewcommand{\arraystretch}{2}
\begin{array}{c|c|c}
d\, & 14 & 16\\ \hline
\zeta_{Y_d}(0)\, &\, \frac{157009}{122594472000}\, &\, -\frac{16215071}{62523180720000} 
\end{array}
\label{eq:zeta0}
\end{align}
and
using these values into the general expression~\eqref{eq:tr-z} 
produces the type-A trace anomalies that match our results 
of Table~\ref{table}. 

More recently, within the AdS/CFT paradigm, it was shown how to directly reproduce the $\zeta$-function for a class 
of conformal operators~\cite{Diaz:2008hy}---see also~\cite{Dowker:2010qy}
for a direct proof that does not use holography.
For the quadratic operator $Y_d$ this amounts to compute the following integral
\begin{align}
\zeta_{Y_d}(0) = \frac{2(-)^{d/2}}{d!} \int_0^1 d\nu \prod_{l=0}^{d/2-1}\big( l^2-\nu^2\big)\,,\quad d>2~,
\end{align}  
which can be easily checked to reproduce~\eqref{eq:zeta0}. 

\section{Conclusion and outlook} 
\label{sec:6} 

Mastering the computation of scattering amplitudes that involve gravitons is an outstanding task that keeps drawing the 
attention of many theoretical physicists---recently, for example, several interesting papers have dealt with the issue 
of soft graviton insertions in scattering 
amplitudes, see for example~\cite{Strominger:2013jfa, Cachazo:2014fwa, Bern:2014vva,Sen:2017nim, Laddha:2017ygw}
or the pedagogical review~\cite{Strominger:2017zoo}.
From the worldline formalism viewpoint, the main difficulty in tackling the computation of graviton scattering amplitudes resides in the presence of derivative interactions in the nonlinear sigma model, that represents the first quantized particle in a generically curved space. In the present paper, following the developments of ref. \cite{Bastianelli:2017wsy},
 we have investigated further the use of an effective linear sigma model to study the one-loop effective action 
 of a scalar field in a maximally-symmetric curved space, and its type-A trace anomaly in particular.
 In the literature, other and certainly more efficient methods to compute type-A
 trace anomalies of conformal QFT's are known---often based on the  $\zeta$-function approach to 
 compute determinants.
 However, unlike those methods, the present approach is much more flexible, allowing for example to compute 
 the off-diagonal parts of the heat kernel and, in general, to give a worldline representation  of the QFT observable 
 that one wishes to study,  see for example the recent use of a worldline representation
 to relate different quantities made in \cite{Giombi:2017txg}.

To extend further the use of the linear sigma model approach, it would be interesting to
prove its validity on arbitrary geometries, a possibility already envisaged in \cite{Guven:1987en},
but whose implementation might be obstructed by backgrounds with less symmetries
than the maximal one.
  
 Considering only spaces with maximal symmetries, a still useful extension would be the
 introduction of worldline fermions, so to be able to consider $N=1$ and $N=2$ supersymmetric 
 generalizations, as needed in the worldline description of spin 1/2 and spin 1 particles.
 An extension to arbitrary $N$  would also allow to study higher spinning particles
on maximally symmetric spaces~\cite{Kuzenko:1995mg, Bastianelli:2008nm}.
 In the nonlinear sigma model approach the regularizations and counterterms for the 
 supersymmetric version of the path integral of a particle in a curved space have 
  been most extensively analyzed at arbitrary $N$ in \cite{Bastianelli:2011cc}. 
 A linear sigma model approach would carry many simplifications and would 
 certainly be welcome. In the case of spin 1/2, one might wish 
 to study from a worldline perspective  the issue of the trace anomaly of a Weyl fermion, 
 where an apparent clash between the results of \cite{Bonora:2014qla, Bonora:2017gzz} 
 and \cite{Bastianelli:2016nuf} has emerged. However, to address that point with worldline methods requires 
mastering the use of a  generic background, as the conflicting result sits in the coefficient of a type-B trace anomaly.

\appendix   
\section{Geometry of maximally symmetric spaces and Riemann normal coordinates}
\label{appA}

Maximally symmetric spaces are those that have a maximal number of isometries,
namely $d(d+1)/2$ for a $d$-dimensional space. 
Their curvature tensors can be expressed in terms of the metric as  
\bea
R_{ijmn} \eqa M^2 (g_{im}g_{jn}-g_{in}g_{jm} ) \\[2mm]
R_{ij} \eqa R_{mi}{}^m{}_j = M^2 (d-1) g_{ij}  \\[2mm]
R \eqa R_{i}{}^i =
M^2 (d-1) d ~,
\eea
where  $M^2$ is a constant which identifies the sectional curvature of the manifold. This constant is positive on a  sphere of radius $a$, where $M^2=1/a^2$, it vanishes for a flat space, and it is negative for a real hyperbolic space. 
This exhausts the list of maximally symmetric spaces. 
For simplicity in the main text we have considered spheres, but here we treat briefly real hyperbolic spaces as well.

In the main text we use Riemann normal coordinates (for details see \cite{Eisenhart:1965, Petrov:1969}, and
\cite{Honerkamp:1971sh, AlvarezGaume:1981hn, Howe:1986vm} for their application to nonlinear sigma models;  
the most accurate and explicit expansion of the metric around the origin that we are aware of  may  
be found in \cite{Bastianelli:2000dw}).
On spheres the sectional curvature is positive, and we can take $M=\frac{1}{a} >0$. It  
is then easy to evaluate recursively all terms in the expansion of the metric \cite{Bastianelli:2001tb}
\be
 g_{ij}(x) =  \delta_{ij} + \sum_{l=1}^{\infty} c_l M^{2l} (-1)^l (x^2)^l P_{ij} 
 = \delta_{ij} + f(x) P_{ij} ~,  \label{A4}
\ee
where $x^i$ denote now the Riemann normal coordinates centered around a point (the origin), 
$P_{ij}$  indicates a projector given by
\be
P_{ij} = \delta_{ij} - \hat x_i \hat x_j ~,  \qquad  \hat x^i = \frac{x^i}{x} ~, \qquad  
x = \sqrt{\vec x^{\, 2}} ~,
\ee
and $c_l$ are coefficients that obey the recursion relation
\be
c_l = \frac{2}{(l+1)(2l+1)} c_{l-1} ~, \qquad c_0=1~,
\ee
and are found to be given by
\be
c_l =  2\frac{4^l}{(2l +2)!} \;.
\ee
The series can be  summed up to give
\be
f(x) = \frac{1-2 (Mx)^2 - \cos(2Mx)}{ 2 (Mx)^2}~,  \label{A5} 
\ee
which was also reproduced in \cite{Bastianelli:2001tb} (there is a misprint  in Eq. (11) of \cite{Bastianelli:2001tb}, where 
a factor $(x^2)^2$  in the denominator should be replaced by $x^2$).  The present manuscript has the correct answer.
Note that the function $f(x)$ does not have poles and it is even in $x$, so that it depends only on 
$x^2=\vec x^{\, 2} =\delta_{ij}x^i x^j$.  Note also that, because of the projector $ P_{ij}$
one has the equality  $x^2=g_{ij}(x)  x^i x^j$.
It is now immediate to compute the inverse metric $ g^{ij}(x)$ and metric determinant   $g(x)$ as
\bea
 g^{ij}(x) \eqa \delta^{ij} + h(x) P^{ij}  \label{A8} \\[2mm]
  g(x)\eqa (1+f(x))^{d-1} ~,
\eea
where
\be
 h(x) = -\frac{f(x)}{1+f(x)} \label{A10}\;.
\ee
We recall again that on the right hand side of these formulae indices are raised and lowered with the 
flat metric $\delta_{ij}$.

For completeness, we discuss the case of real hyperbolic spaces as well. Now the sectional curvature is negative,
$M^2<0$. It can be obtained form the previous case by the analytic continuation
$M \to i |M| $, with the imaginary unit $i$ giving rise to the negative sign of the sectional curvature, and 
$|M| =\sqrt{- M^2}$. 
Performing this analytic continuation in \eqref{A4} we find that in the sum the minus signs from $(-1)^l$ get canceled
\be
 g_{ij}(x) =  \delta_{ij} + \sum_{l=1}^{\infty} c_l |M|^{2l} (x^2)^l P_{ij} 
 = \delta_{ij} + f(x) P_{ij}   ~,
\ee
and the sum now converges to the function
\be
f(x) = \frac{-1-2 (|M|x)^2 + \cosh(2|M|x)}{ 2 (|M| x)^2} \;.
\ee

Finally, the function $f(x)$ vanishes in the flat space case, where Riemann normal coordinates are just the standard
cartesian coordinates. It may also be obtained as a smooth limit of the curved cases, as $f(x)\to 0$ for $M\to 0$.

\section{Perturbative calculations}
\label{appB}

We describe here the perturbative calculations needed to identify the corrections of order $\beta^7$ and $\beta^8$
to the transition amplitude \eqref{peta}. Lower orders have been computed in~\cite{Bastianelli:2017wsy}.

At order $\beta^7$ we need to evaluate 
\be
- \la S_{12} \ra   +
\la S_8 S_2\ra_c  +
\la S_6 S_4\ra_c  -\frac12
\la S_2^2 S_4\ra_c 
\ee
where $\la ...\ra_c$ indicates connected correlation functions. 
In reporting our intermediate results we set $M=1$ (sphere of unit radius),
use the abbreviation $\Delta(\tau_1,\tau_2)\equiv\Delta_{12}$
for the propagator, and indicate the contributions from topologically distinct Wick contractions
that give rise to different powers of the dimension $d$, so to help for a verification of our intermediate results.

The different contributions are as follows
\be
-\la S_{12}\ra = -\beta^7 k_{12}
\underbrace{\Big ( d^6 + 30 d^5 + 340d^4 +  1800 d^3 +
4384 d^2 +3840 d \Big )}_{d(d+2)(d+4)(d+6)(d+8)(d+10)}
\underbrace{\int_0^1 \!\! d\tau_1 \, \Delta_{11}^6}_{\frac{1}{12012}} 
\ee

\be  
\la S_8 S_2\ra_c  = -\beta^7 k_{8}k_{2}
\underbrace{\Big ( 8 d^4 + 96 d^3 + 352d^2 +  384d \Big )}_{8 d(d+2)(d+4)(d+6)} 
\underbrace{\int_0^1 \!\!  d\tau_1\int_0^1 \!\!  d\tau_2 \, \Delta_{12}^2 \Delta^3_{11}}_{-\frac{1}{8316}} 
\ee

\bea
\la S_6 S_4\ra_c  \eqa -\beta^7 k_{6}k_{4}
\left ( \underbrace{\Big ( 12 d^4 + 96 d^3 + 240 d^2 +192 d\Big )}_{12 d (d+2)^2 (d+4)}  
\underbrace{
\int_0^1 \!\!  d\tau_1\int_0^1 \!\!  d\tau_2 \,\ \Delta_{11}^2 \Delta_{12}^2 \Delta_{22}}_{-\frac{2}{17325}}  \right .
\ccr
&+& \left . 
\underbrace{\Big ( 24 d^3 + 144 d^2 +192 d\Big )}_{24d(d+2)(d+4)} 
\underbrace{
\int_0^1 \!\!  d\tau_1\int_0^1 \!\!  d\tau_2 \,\ \Delta_{11} \Delta_{12}^4}_{-\frac{1}{13860}} 
\right )
\eea

\be
-\frac12\la S_2^2 S_4\ra_c =  -\beta^7 k_2^2 k_4\,  4 d(d+2) \left (
2 \underbrace{
 \int \!\! \!\!\int \!\! \!\!\int
\,  \Delta_{12} \Delta_{13} \Delta_{23}  \Delta_{33}}_{\frac{13}{56700}} 
+ 
\underbrace{
\int \!\! \!\!\int \!\! \!\!\int
\,  \Delta^2_{13} \Delta^2_{23} }_{\frac{1}{5670}} 
\right ) ~.
\ee

At order $\beta^8$ we need  instead
\be
-\la S_{14}\ra  
+\la S_{10} S_2\ra_c 
+\la S_8 S_4\ra_c 
+\frac12 \la S_6^2 \ra_c 
-\frac12 \la S_4^2 S_2\ra_c 
-\frac12 \la S_6 S^2_2\ra_c 
+\frac{1}{4!} \la S_2^4\ra_c 
\ee
and the different contributions are now as follows
\bea
-\la S_{14}\ra =
 \beta^8 k_{14}
\underbrace{\Big ( d^7 + 42 d^6+ 700d^5+5880 d^4 +25984 d^3+56448 d^2 +46080 d
\Big )}_{d(d+2)(d+4)(d+6)(d+8)(d+10)(d+12)}
\underbrace{\int_0^1 \!\! d\tau_1 \, \Delta_{11}^7}_{-\frac{1}{51480}} \ccr
\eea
\be
\la S_{10} S_2\ra_c  = \beta^8 k_{10}k_{2}
 \underbrace{\Big ( 10d^5 +200 d^4 + 1400 d^3 + 4000 d^2 +3840 d\Big )}_{10 d (d+2) (d+4)(d+6) (d+8)}  
\underbrace{
\int_0^1 \!\!  d\tau_1\int_0^1 \!\!  d\tau_2 \,\ \Delta_{11}^4 \Delta_{12}^2 }_{\frac{1}{36036}}  
\ee

\bea
\la S_8 S_4\ra_c  \eqa  
\beta^8 k_{8}k_{4}
\left (  
\underbrace{\Big (  16 d^5+ 224 d^4 + 1088 d^3 + 2176 d^2 + 1536 d
 \Big )}_{16 d (d+2)^2 (d+4)(d+6)}  
 \underbrace{
\int_0^1 \!\!  d\tau_1\int_0^1 \!\!  d\tau_2 \,\ \Delta_{11}^3 \Delta_{12}^2 \Delta_{22}}_{\frac{19}{720720}}  
 \right .
\ccr
&+& \left . 
\underbrace{\Big (48d^4+576 d^3+2112d^2+ 2304 d\Big )}_{48d(d+2)(d+4)(d+6)} 
\underbrace{
\int_0^1 \!\!  d\tau_1\int_0^1 \!\!  d\tau_2 \,\ \Delta^2_{11} \Delta_{12}^4}_{\frac{1}{60060}} 
\right )
\eea

\bea
\frac12 \la S_6^2 \ra_c \eqa \frac12 \beta^8 k_6^2
\left (  
\underbrace{\Big ( 18  d^3 (d+2)^2 + 144d^2 (d+2)^2 +288 d (d+2)^2 \Big )}_{18 d (d+2)^2 (d+4)^2}  
 \underbrace{ \int \!\! \!\!\int
 \Delta_{12}^2 \Delta_{11}^2 \Delta_{22}^2}_{\frac{491}{18918900}}  
 \right .
\ccr
&+& 
\underbrace{\Big ( 72  d^3 (d+2) + 576 d^2 (d+2) + 1152 d (d+2) \Big )}_{72 d (d+2) (d+4)^2}  
 \underbrace{ \int \!\! \!\!\int  \Delta_{12}^4 \Delta_{11} \Delta_{22}}_{\frac{25}{1513512}}  
\ccr
&+&  \left .
\underbrace{\Big ( 48  d^3  + 288 d^2 + 384 d \Big )}_{48 d (d+2) (d+4)}  
 \underbrace{ \int \!\! \!\!\int  \Delta_{12}^6}_{\frac{1}{84084}}  
\right )
\eea

\bea
-\frac12 \la S_4^2 S_2\ra_c \eqa  
 \frac12 \beta^8 k_4^2 k_2 
\left (   
32 d (d+2)^2 \left (
\underbrace{ \int \!\! \!\!\int \!\! \!\!\int  \Delta^2_{12}\Delta^2_{23}\Delta_{33}}_{-\frac{2}{51975}}
+
\underbrace{ \int \!\! \!\!\int \!\! \!\!\int  \Delta_{12}\Delta_{13}\Delta_{23}\Delta_{22}\Delta_{33}}_{-\frac{83}{1663200}}
 \right )
 \right .
\ccr
&+& \left .
64 d (d+2) 
\underbrace{ \int \!\! \!\!\int \!\! \!\!\int  \Delta_{12}\Delta_{13}\Delta^3_{23}}_{-\frac{1}{34650}}
\right )
\eea

\bea
-\frac12 \la S_6 S^2_2\ra_c \eqa  
 \frac12 \beta^8 k_6 k_2^2\, 
24 d (d+2)(d+4) \left (
\underbrace{ \int \!\! \!\!\int \!\! \!\!\int  \Delta^2_{11}\Delta_{12}\Delta_{13}\Delta_{23}}_{-\frac{8}{155925}}
+
\underbrace{ \int \!\! \!\!\int \!\! \!\!\int  \Delta_{11}\Delta^2_{12}\Delta^2_{13}}_{-\frac{1}{24948}}
 \right )\ccr
\eea

\be
\frac{1}{4!} \la S_2^4\ra_c  = \frac{1}{4!} \beta^8 k_2^4\, 48 d 
\underbrace{ \int \!\! \!\!\int \!\! \!\!\int   \!\! \!\!\int   
\Delta_{12}\Delta_{23}\Delta_{34}\Delta_{41}}_{\frac{1}{9450}} \;.
\ee

We may now insert the values of the coupling constants
{\allowdisplaybreaks
\bea
k_2 \eqa (d-1)(d-3) \frac{1}{120} 
\ccr[2mm]
k_4 \eqa (d-1)(d-3) \frac{1}{756} 
\ccr[2mm]
k_6 \eqa (d-1)(d-3) \frac{1}{5400}
\ccr[2mm]
k_8 \eqa (d-1)(d-3) \frac{1}{41580}
\ccr[2mm]
k_{10} \eqa (d-1)(d-3) \frac{691}{232186500}
\ccr[2mm]
k_{12} \eqa (d-1)(d-3)  \frac{1}{2806650}
\ccr[2mm]
k_{14} \eqa (d-1)(d-3) \frac{3617}{86837751000}
\eea
}
found from \eqref{K-coeff}, reintroduce the correct power of $M$,
and add all terms to find the final answer reported in~\eqref{final-1} or equivalently in~\eqref{final-2}.


\end{document}